\documentclass[aip,apl,reprint]{revtex4-1}

\usepackage{graphicx}
\usepackage{dcolumn}
\usepackage{bm}
\begin{document}

\preprint{APS/123-QED}

\title{Phase noise analysis of mutually synchronized spin Hall nano-oscillators}

\author{Artem Litvinenko}
\thanks{These authors contributed equally.}
\affiliation{Applied Spintronics Group, Department of Physics, University of Gothenburg, 412 96 Gothenburg, Sweden}%

\author{Akash Kumar}
\thanks{These authors contributed equally.}
\affiliation{Applied Spintronics Group, Department of Physics, University of Gothenburg, 412 96 Gothenburg, Sweden}%
\affiliation{Center for Science and Innovation in Spintronics, Tohoku University, 2-1-1 Katahira, Aoba-ku, Sendai 980-8577 Japan}

\author{Mona Rajabali}
\affiliation{NanOsc AB, Kista, Sweden}%

\author{Ahmad A. Awad}
\affiliation{Applied Spintronics Group, Department of Physics, University of Gothenburg, 412 96 Gothenburg, Sweden}%
\affiliation{Center for Science and Innovation in Spintronics, Tohoku University, 2-1-1 Katahira, Aoba-ku, Sendai 980-8577 Japan}
\affiliation{Research Institute of Electrical Communication, Tohoku University, 2-1-1 Katahira, Aoba-ku, Sendai 980-8577 Japan}

\author{Roman Khymyn}
\affiliation{Applied Spintronics Group, Department of Physics, University of Gothenburg, 412 96 Gothenburg, Sweden}%

\author{Johan \AA kerman}%
 \email{johan.akerman@physics.gu.se}
\affiliation{Applied Spintronics Group, Department of Physics, University of Gothenburg, 412 96 Gothenburg, Sweden}
\affiliation{Center for Science and Innovation in Spintronics, Tohoku University, 2-1-1 Katahira, Aoba-ku, Sendai 980-8577 Japan}
\affiliation{Research Institute of Electrical Communication, Tohoku University, 2-1-1 Katahira, Aoba-ku, Sendai 980-8577 Japan}

\date{\today}

\begin{abstract}
The reduction of phase noise in electronic systems is of utmost importance in modern communication and signal processing applications and requires an understanding of the underlying physical processes. Here, we systematically study the phase noise in mutually synchronized chains of nano-constriction spin Hall nano-oscillators (SHNOs). We find that longer chains have improved phase noise figures at low offset frequencies (1/$f$ noise), where chains of two and ten mutually synchronized SHNOs have 2.8 and 6.2 dB lower phase noise than single SHNOs. This is close to the theoretical values of 3 and 10 dB, and the deviation is ascribed to process variations between nano-constrictions. However, at higher offset frequencies (thermal noise), the phase noise unexpectedly increases with chain length, which we ascribe to process variations, a higher operating temperature in the long chains at the same drive current and phase delays in the coupling between nano-constrictions.
\end{abstract}

\maketitle


Spin transfer and spin-orbit torque provide means to drive nanomagnetic systems into current tunable high-frequency precession~\cite{slonczewski1996current,tsoi2000generation,kiselev2003microwave}. The resulting microwave voltage signal can be used for communication applications~\cite{rippard2004,bonetti2009spin,pufall2005frequency,sharma2015modulation,litvinenko2021analog} and spectral analysis\cite{litvinenko2020ultrafast, litvinenko2022ultrafast}, where the small footprint, ready integration with CMOS technology, and wide frequency tunability make these oscillators particularly interesting. While STNOs, comprising ferromagnetic/non-magnetic/ferromagnetic structures, requires a somewhat complex fabrication process due to the current flowing out-of-plane, spin-orbit torque-driven spin Hall nano-oscillators (SHNOs) utilize in-plane currents in simple ferromagnetic/heavy metal bilayer systems~\cite{VEDemidov2012,demidov2014nanoconstriction,durrenfeld2017nanoscale,awad2020width,behera2022energy,zahedinejad2018cmos,fulara2019spin,fulara2020giant,haidar2019single,gonzalez2022voltage}, where heavy metals (e.g. Pt~\cite{mazraati2018improving}, Ta~\cite{tiwari2017antidamping,kumar2018large}, and W~\cite{bansal2018large,mazraati2016low,behera2022energy}) produce pure spin currents through the spin Hall effect~\cite{sinova2015rmp}. 
The simple geometry and in-plane current flow allow ease of fabrication, direct optical access, and the ability to synchronize multiple oscillators in chains and two-dimensional arrays~\cite{awad2017long,zahedinejad2018cmos,kumar2023robust}, making such SHNOs promising candidates for emerging spintronic applications including Ising machines~\cite{csaba2020coupled, houshang2022SHNOIM, albertsson2021ultrafastSHNOIM} and neuromorphic computing~\cite{torrejon2017neuromorphic,romera2018vowel,zahedinejad2020two,Zahedinejad2022natmat}. 
For communication applications, the phase noise plays a crucial role, as it directly determines the performance of the system. To evaluate the potential of nano-oscillators for conventional signal processing applications, it is hence essential to characterize their phase noise performance~\cite{schneider2009temperature,SilvaPhaseNoiseTheory2010,keller2010nonwhite,eklund2014dependence,sharma2014mode,lebrun2015understanding,wittrock2019low}, understand its physical origin, and suggest methods~\cite{wittrock2021stabilization} for its improvement. 

\begin{figure}[hbt!]
\includegraphics[width=7.8cm]{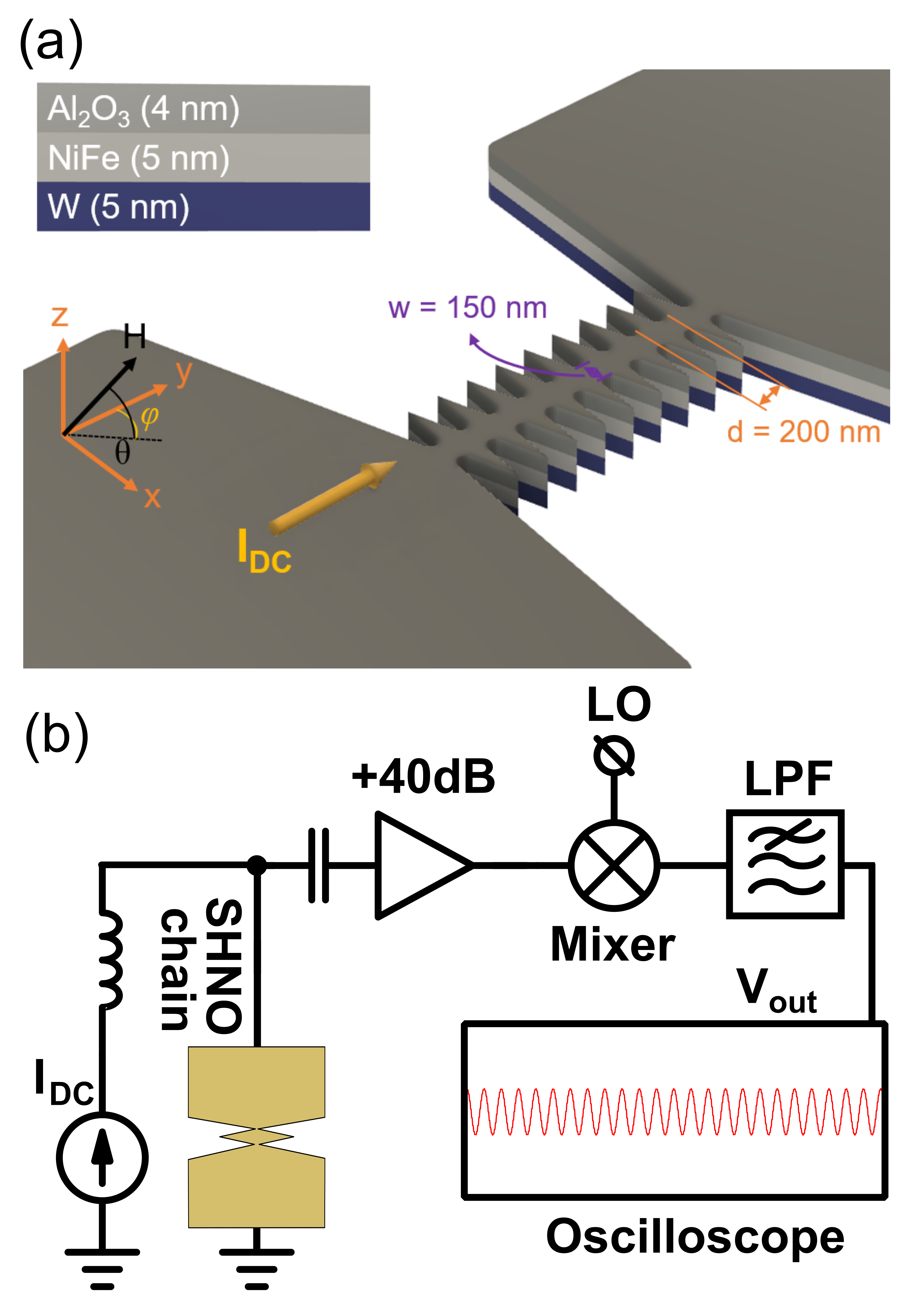}%
\caption{\label{Fig1} (a) Schematic of a SHNO chain with 10 nano-constrictions in series. (b) The phase noise measurement setup, where 
LO is a local oscillator tuned around 18 GHz to down-convert the SHNO signal to 10 MHz. The mixer is a ZMDB-44H-K+ double-balanced mixer, and the LFP is a lumped-element lowpass filter with a cutoff frequency of 30 MHz and stopband attenuation of 40dB. The sampling rate of the oscilloscope is 50 MS/s.}
\end{figure}

\begin{figure*}[t!]
\includegraphics[width=17.8cm]{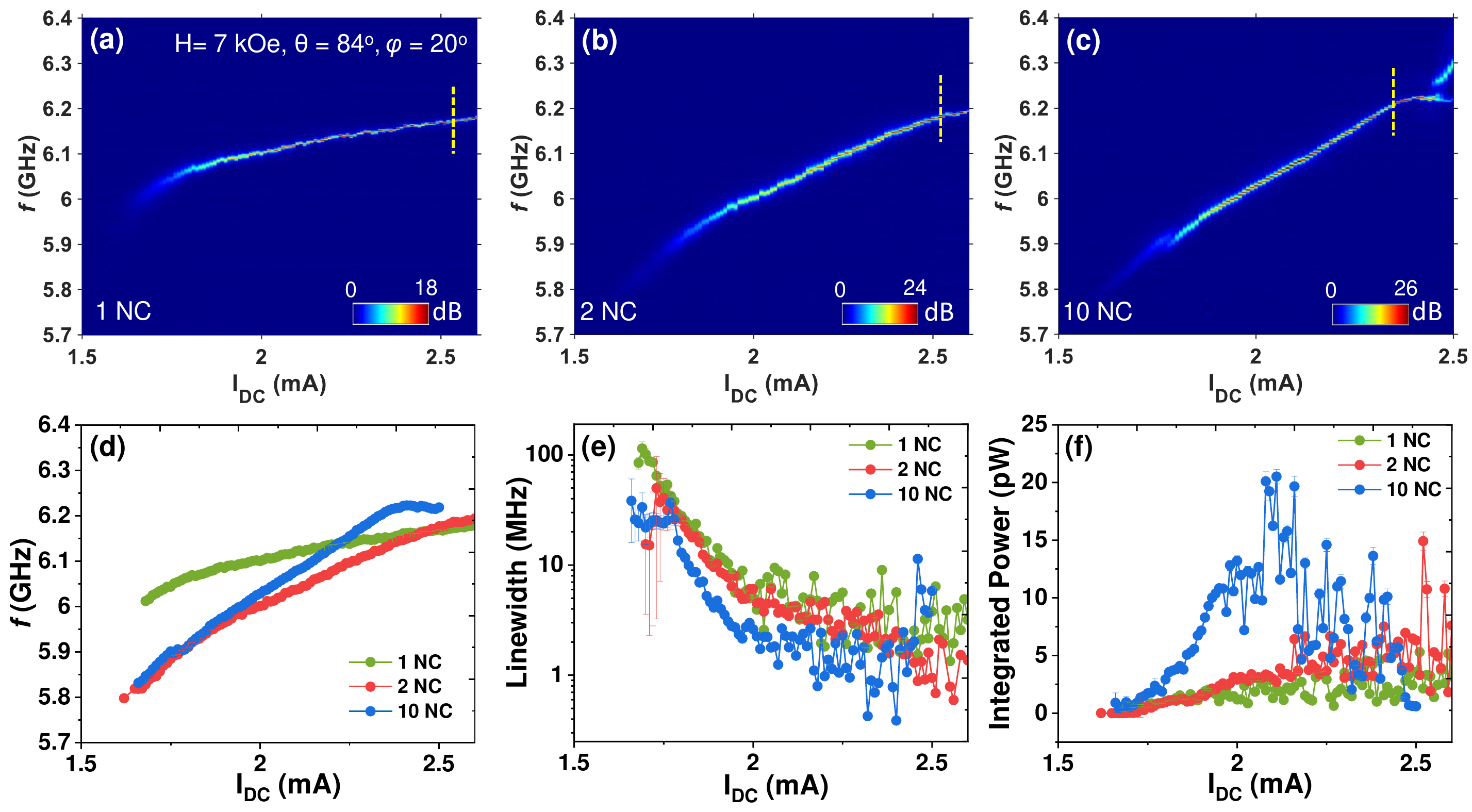}%
\caption{\label{Fig2} Free running properties of single and mutually synchronized 2 NC and 10 NC SHNO chains. Power spectral density (PSD) of the auto-oscillation for (a) single NC, (b) 2NC, and (c) 10 NC, respectively. Extracted (d) auto-oscillation frequency, (e) linewidth, and (f) integrated power. The dashed yellow line represents the current used during phase noise measurements.}
\end{figure*}

Here, we perform a comprehensive analysis of phase noise in single NC SHNOs as well as short (2 NCs) and longer (10 NCs) chains of mutually synchronized NC SHNOs and demonstrate that it can be significantly reduced compared to single SHNOs. We find that in the case of two NC SHNOs the mutual synchronization leads to an improvement of 2.8~dB in the $1/f$ flicker frequency phase noise, which is very close to the theoretical prediction of 3~dB. In the longer chain of 10 NCs, the $1/f$ noise improves by 6.2~dB, which is substantial but further (3.8~dB) from the theoretical expectation of 10~dB. We argue that this deviation originates from process variations between individual NCs, since the theoretical value assumes identical intrinsic frequencies of all oscillators. Somewhat unexpectedly, the white (thermal) frequency phase noise at higher noise frequencies is found to increase with chain length, being 2.1~dB worse for two NCs and 3.1~dB worse for 10 NCs, compared to single SHNOs. In addition to process variations, the longer chains also operate at higher temperature, due to the higher power dissipation, which may further increase the phase noise, in particular in the thermal region. As the measured linewidth improves substantially with chain length, we conclude that it is governed primarily by the $1/f$ noise. 

\begin{figure*}[t!]
\includegraphics[width=17.8cm]{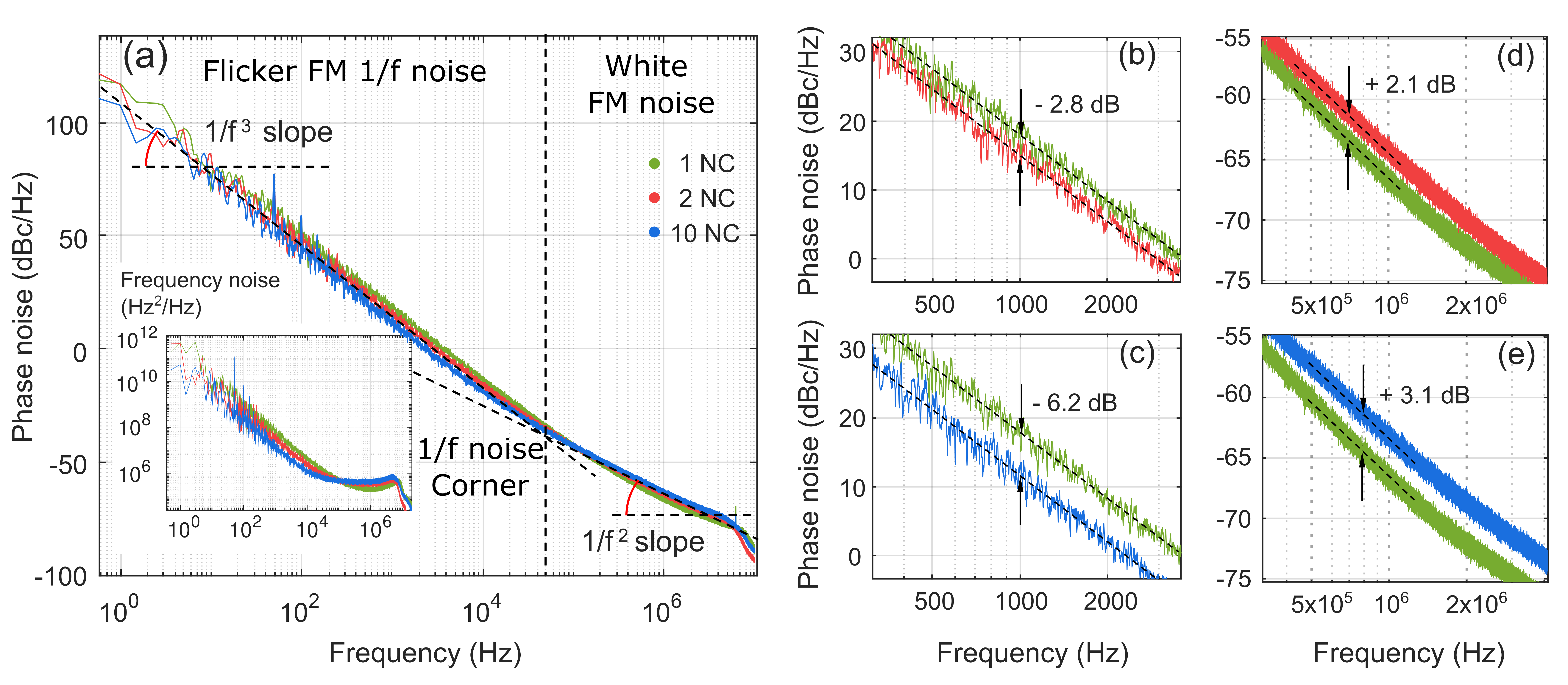}%
\caption{\label{Fig3} Phase noise spectrum plot for a single and mutually synchronized 2 NC and 10 NC SHNOs in a chain. The dashed vertical line represents the $1/f$--corner frequency of 50 kHz and separates regions with flicker frequency noise and white frequency phase noise. The steep reduction in phase noise above 6 MHz is associated with the applied bandpass filter used to improve SNR.}
\end{figure*}


Single NC SHNOs and chains were fabricated from DC/RF magnetron sputtered W(5 nm)/NiFe(5 nm)/Al$_{2}$O$_{3}$(4 nm) stacks. 
The large spin Hall angle of W ($|\theta_{SH}|>$0.44) reduces the threshold current~\cite{mazraati2016low,behera2022energy} and the anisotropic magnetoresistance (AMR) of NiFe (0.65~$\%$) provides a reasonable output power~\cite{mazraati2018improving,rajabali2023injection}. The devices were patterned into 150 nm NCs with 200 nm center-to-center separation (in chains) using e-beam lithography followed by Ar-ion etching (for details see \emph{e.g.}~\cite{kumar2022fabrication}). Figure~\ref{Fig1}a shows a schematic of a 10 NC chain.

Phase noise measurements were performed at fixed current and magnetic field and analyzed using a Hilbert transform technique \cite{litvinenko2021analog}. Analysis of close-in phase noise at low offset frequencies of sub-Hz range requires that the experimental time traces are accumulated over seconds time scales, which would require Terabytes of data to be processed with direct signal sampling at a 40 GS/s rate. In order to reduce the processed amount of data for such a long time series, we performed SNHO signal down-sampling using a frequency mixer as shown in Fig.~\ref{Fig1}(b). The NC SHNO signal is downconverted to 10 MHz by adjusting the local oscillator (LO) frequency and captured with a real-time oscilloscope at a low sampling rate of 50 MS/s. We process the captured signal in several steps. First, the captured SHNOs signal gets filtered with an FIR digital band-pass filter with a central frequency of 10~MHz, bandwidth of 12~MHz and stopband attenuation of 60~dB to improve the signal-to-noise ratio (SNR) so that the amplitude of the SHNO signal is sufficiently higher than the RMS amplitude of the thermal noise floor. Note that even though the bandpass filter removes the thermal noise floor it still allows the analysis of the close-in phase noise of the signal. Then, the instantaneous phase gets extracted with a Hilbert transform~\cite{rabiner1975theory,boashash1992estimating,feldman2011hilbert} of signal time traces. At the next step, the instantaneous phase signal is detrended. Finally, a power spectrum density is calculated from the detrended instantaneous phase signal using FFT.


The free-running auto-oscillations with varying DC current (I$_{DC}$) are shown in Fig.~\ref{Fig2}(a,b and C) for single (1 NC), two (2 NC) and ten (10 NC) mutually synchronized SHNOs, respectively. Figure~\ref{Fig2}d summarizes their operating frequency, where it can be observed that mutual synchronization of SHNOs leads to higher frequency tunability. This could be understood as an absolute increase in their magneto-dynamical region. Figure~\ref{Fig2}e and \ref{Fig2}f shows the linewidth and integrated output power for the oscillator chains. It is clear from the observed parameters that mutual synchronization leads to larger output power and lower linewidth for a larger number of synchronized oscillators in a chain, consistent with earlier work~\cite{awad2017long,kumar2023robust}.  

The results of phase noise measurement are presented in Fig.~\ref{Fig3}(a-e) for a single and mutually synchronized 2 NC and 10 NC oscillators in a chain. Phase noise measurements are performed at I$_{DC}$ = 2.53 mA, 2.53 mA and 2.35 mA (also indicated by the yellow dashed line in Fig.~\ref{Fig2}a-c) for single and mutually synchronized 2 NC and 10 NC oscillators, respectively. As can be seen from Fig.~\ref{Fig3}(a) all devices demonstrate regions with $1/f$ and white (thermal) frequency 
phase noise. Interestingly, in our SHNOs, the $1/f$ phase noise corner appears at a much lower offset frequency of 50 kHz compared to the GHz frequencies for MTJ STNOs~\cite{sharma2014mode}. A lower value of the $1/f$ corner leads to an improved linewidth at laboratory time scales as the $1/f$ noise has a steep slope and a much higher contribution to the integrated power of phase noise.  

A single SHNO exhibits a phase noise of $0$ , $-17$, and $-67$ dBc/Hz at the offset frequencies 100 Hz, 10 kHz, and 1 MHz, respectively. In Fig.~\ref{Fig3}(b) it can be seen that two mutually synchronized SHNOs demonstrate a 2.8~dB improvement in the $1/f$ region, which is in good agreement with the theoretical expectation of 3~dB improvement for each doubling in the number of synchronized identical oscillators~\cite{ChangPhaseNoise1997, pikovsky_rosenblum_kurths_2001, zhang2015synchronization}. In case of 10 NC SHNOs, we expect to see a 10 dB improvement, but the experiment only showed a reduction of 6.2 dB. This may be attributed to process variations in the NC width, which lead to variations in the intrinsic frequency of the nano-constrictions in the chain. Process variations naturally become more noticeable in longer chains as the probability to find $N$ identical oscillators decreases rapidly with $N$. Another factor could be attributed to the geometry of the chain and its associated thermal effects. A chain of two identical nano-constrictions will retain the same zero difference in their relative frequencies even when the temperature and its gradient increase. 
However, in the case of more than two coupled NC-SHNOs, the inner and outer NCs  will heat up differently, leading to a varying intrinsic frequency as a function of position in the chain. 

Unexpectedly, in the region of white frequency phase noise, the 2 NC and 10 NC SHNO, instead of an improvement, show an \emph{increase} of the phase noise by 2.1 and 3.1 dB, respectively, as compared to a single NC SHNO (see Fig.~\ref{Fig3}(d)). A possible explanation could be that process variations affect thermal noise much more than the $1/f$ noise. From Fig.~\ref{Fig2}(d) we can deduce from the increase of the frequency variation with current that nonlinearity of NC SHNO chains increases with the number of oscillators. It may lead to a sufficient shift in the corner of white frequency phase noise. Additionally, the temperature of the NC SHNO is higher for longer chains which contributes to the region of up-converted thermal noise. From the inset of Fig.~\ref{Fig3}(a) where we plot frequency noise, it is more evident that for 2 and 10 NC the level of white frequency noise, which corresponds to the flicker frequency type of phase noise, increases with chain length.

\begin{figure}[t!]
\includegraphics[width=8.2cm]{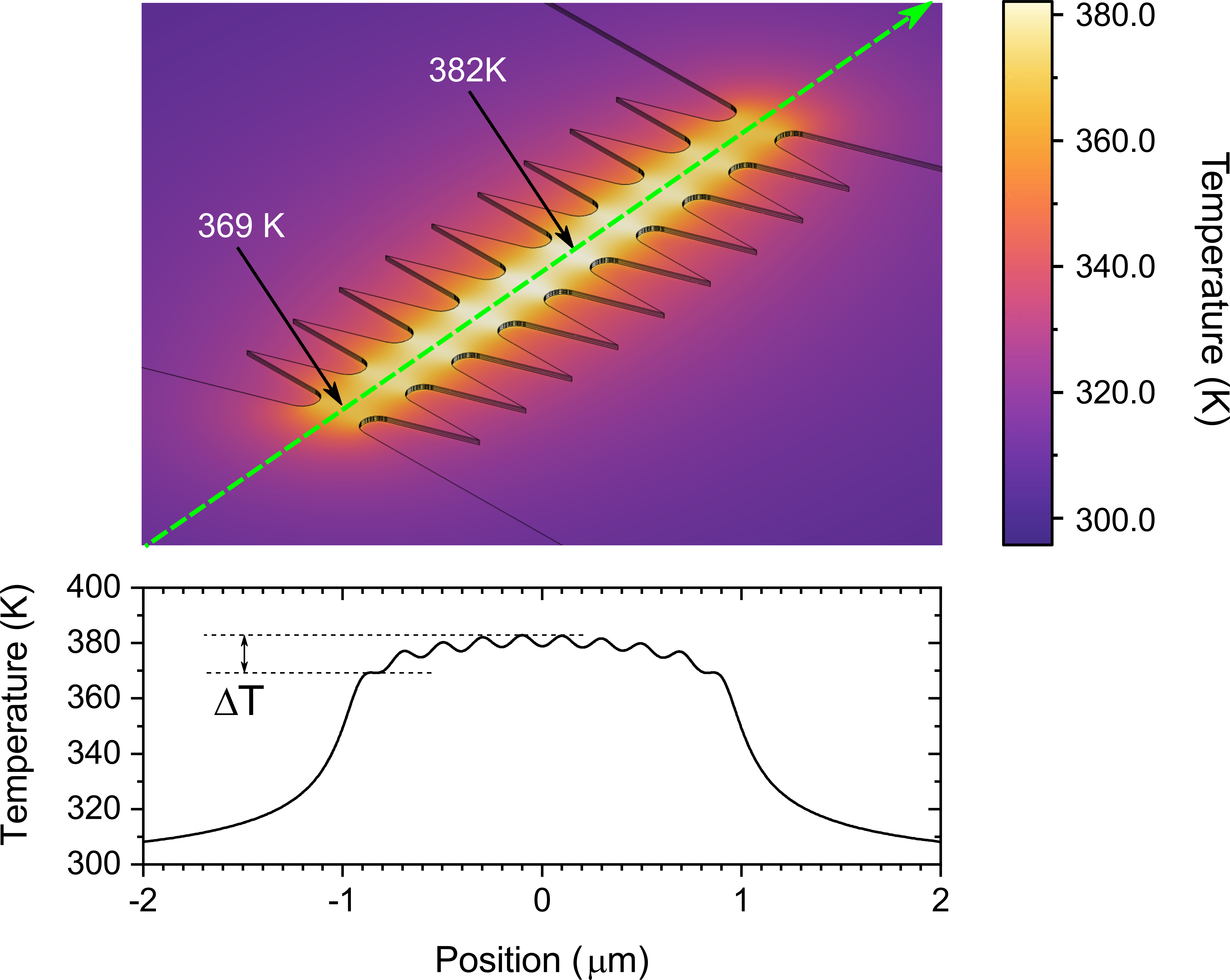}%
\caption{\label{Fig4} COMSOL simulation of 10 NC SHNOs in a chain. Top panel: a thermal map for an applied DC current of 2.35~mA. Bottom panel: a temperature profile along x-axis depicted as a dashed green line in top panel. The temparature difference $\Delta$T between the cental and the edge NC SHNOs is 13K.}
\end{figure}

To understand the extent of the temperature gradient in long chains, we performed COMSOL simulations of a 10 NC SHNO chain. We used the COMSOL moduel Electric Currents (ec) to simulate the current density variation in the nanoconstrictions together with the Heat Transfer in Solids (ht) module. Multiphysics simulations were performed using the Electromagnetic Heating (emh1) module. In our simulation, we took into account the 2~nm silicon oxide layer on top of the silicon wafer which has a significantly lower thermal conductivity of 1.4~W/(m*K). The base silicon wafer has a thermal conductivity of 34~W/(m*K). The simulations are performed using the measured resistivity for the thin films i.e. W (300 $\mu \Omega$-cm) and NiFe (40 $\mu \Omega$-cm). In order to reduce the simulation time and resources we simulate a limited chip area of 1.5x1.5x0.5~mm. Temperature boundary conditions of 293.15~K are applied at the edges of the simulated area. The top panel in Fig.~\ref{Fig4} shows a thermal map for an applied DC current of 2.35~mA flowing through the chain. In order to visualize the temperature gradients we have plotted a temperature profile along x-axis in bottom panel of Fig.~\ref{Fig4}. It can be seen that the temperature gradient exponentially increases to the edge of an array. The temperature deviation $\Delta$T between the central and the outer NC SHNOs is 13~K. In our previous studies~\cite{muralidhar2022optothermal}, we have experimentally observed a large change in operating frequency due to thermal effects. In our present work, we estimate that the temperature gradient contributes a 20~MHz change in intrinsic frequency between the oscillators. However, since a deviation of 20~MHz unequivocally falls within the broad locking range of SHNOs~\cite{awad2017long} it cannot be the main factor of the sufficient increase in phase noise.
Another reason that can lead to an increase of phase noise in mutually synchronized chains of oscillators with primarily nearest-neighboor coupling is the phase delay in the coupling. In the paper~\cite{XianhePhaseNoise2007} it has been shown that the total phase noise can sufficiently increase in a chain of oscillators with nearest-neighboor coupling. Since NC SHNO chains demonstrate positive nonlinearity the coupling between oscillators most likely happens through propagating spinwaves which may lead to a large delay. The phase delay of the coupling between NC SHNOs has to be explored further in order to fully understand its contribution to the phase noise increase in both flicker frequency and white frequency regions of the phase noise.

In summary, we have analyzed the phase noise for single, double and ten nano-constriction SHNOs. Two mutually synchronized SHNOs demonstrate a 2.8 ~dB reduction in phase noise, which corresponds well to the theoretical estimation of 3~dB. The longer chains of 10 nano-constrictions demonstrate an improvement of 6.2~dB, which is further from the theoretical value of 10~dB and can be associated with several factors such as \emph{i}) process variation of the nano-constrictions, \emph{ii}) temperature gradients within the chain making the NC SHNOs nonidentical and increasing the overall temperature, and \emph{iii}) phase delays in the coupling between nano-constrictions, which may  lead to decoherence in the chain and elevated noise levels. Further phase noise measurements and analysis will be required for a more complete understanding of these different mechanisms and ways to mitigate their impact. 

\section*{Acknowledgement}
This work was partially supported by the Horizon 2020 Research and Innovation Program (ERC Advanced Grant No. 835068 “TOPSPIN” and Grant No. 899559 “SpinAge”,DOI 10.3030/899559) and 
the Swedish Research Council (VR; Dnr 2016-05980).

\section*{Author Declarations}
\subsection{Conflict of Interest}
The authors have no conflicts to disclose.


\section*{Data Availability}
The data that support the findings of this study are available
from the corresponding authors upon reasonable request.

\bibliography{Main}

\end{document}